\documentclass[a4paper,12pt]{article}

\usepackage{epsfig}

\usepackage{amssymb}
\usepackage{amsfonts}

\newskip\humongous \humongous=0pt plus 1000pt minus 1000pt

\newif\ifdtup

\catcode`\@=11

\@addtoreset{equation}{section}
\def\theequation{\thesection.\arabic{equation}}

\def\@normalsize{\@setsize\normalsize{15pt}\xiipt\@xiipt
\abovedisplayskip 14pt plus3pt minus3pt%
\belowdisplayskip \abovedisplayskip
\abovedisplayshortskip \z@ plus3pt%
\belowdisplayshortskip 7pt plus3.5pt minus0pt}

\def\small{\@setsize\small{13.6pt}\xipt\@xipt
\abovedisplayskip 13pt plus3pt minus3pt%
\belowdisplayskip \abovedisplayskip
\abovedisplayshortskip \z@ plus3pt%
\belowdisplayshortskip 7pt plus3.5pt minus0pt
\def\@listi{\parsep 4.5pt plus 2pt minus 1pt
      \itemsep \parsep
      \topsep 9pt plus 3pt minus 3pt}}

\relax

\catcode`@=12

\catcode`\@=11

\def\section{\@startsection{section}{1}{\z@}{3.5ex plus 1ex minus
    .2ex}{2.3ex plus .2ex}{\large\bf}}

\def\thesection{\arabic{section}}
\def\thesubsection{\arabic{section}.\arabic{subsection}}

\def\appendix{\setcounter{section}{0}
  \def\thesection{Appendix \Alph{section}}
  \def\thesubsection{\Alph{section}.\arabic{subsection}}
  \def\theequation{\Alph{section}.\arabic{equation}}}

\def\SymBoxes#1#2#3#4{\newdimen\un@t \un@t#3%
\raisebox{#1}{\rule{#2\un@t}{#4}\hskip-#2\un@t
\@tempdimb\un@t \advance\@tempdimb by-#4\@tempcntb#2\relax%
\@whilenum{\@tempcntb>0}\do{
\rule{#4}{\un@t}\hskip\@tempdimb \advance\@tempcntb by\m@ne}%
\hskip-#2\un@t \rule[\un@t]{#2\un@t}{#4}%
\rule[\un@t]{#4}{#4}\hskip-#4
\rule{#4}{\un@t}}\hskip-#4}                

\begin{document}


\newcommand{\dd}{\textrm{d}}

\newcommand{\beq}{\begin{equation}}
\newcommand{\eeq}{\end{equation}}
\newcommand{\bea}{\begin{eqnarray}}
\newcommand{\eea}{\end{eqnarray}}
\newcommand{\beas}{\begin{eqnarray*}}
\newcommand{\eeas}{\end{eqnarray*}}
\newcommand{\defi}{\stackrel{\rm def}{=}}
\newcommand{\non}{\nonumber}
\newcommand{\bquo}{\begin{quote}}
\newcommand{\enqu}{\end{quote}}
\renewcommand{\(}{\begin{equation}}
\renewcommand{\)}{\end{equation}}
\def\de{\partial}
\def\Om{\ensuremath{\Omega}}
\def\Tr{ \hbox{\rm Tr}}
\def\H{ \hbox{\rm H}}
\def\HE{ \hbox{$\rm H^{even}$}}
\def\HO{ \hbox{$\rm H^{odd}$}}
\def\HEO{ \hbox{$\rm H^{even/odd}$}}
\def\HOE{ \hbox{$\rm H^{odd/even}$}}
\def\HHO{ \hbox{$\rm H_H^{odd}$}}
\def\HHEO{ \hbox{$\rm H_H^{even/odd}$}}
\def\HHOE{ \hbox{$\rm H_H^{odd/even}$}}
\def\K{ \hbox{\rm K}}
\def\Im{ \hbox{\rm Im}}
\def\Ker{ \hbox{\rm Ker}}
\def\const{\hbox {\rm const.}}
\def\o{\over}
\def\im{\hbox{\rm Im}}
\def\re{\hbox{\rm Re}}
\def\bra{\langle}\def\ket{\rangle}
\def\Arg{\hbox {\rm Arg}}
\def\exo{\hbox {\rm exp}}
\def\diag{\hbox{\rm diag}}
\def\longvert{{\rule[-2mm]{0.1mm}{7mm}}\,}
\def\a{\alpha}
\def\dag{{}^{\dagger}}
\def\tq{{\widetilde q}}
\def\p{{}^{\prime}}
\def\W{W}
\def\N{{\cal N}}
\def\hsp{,\hspace{.7cm}}
\def\bo{\ensuremath{\hat{b}_1}}
\def\bfo{\ensuremath{\hat{b}_4}}
\def\co{\ensuremath{\hat{c}_1}}
\def\cfo{\ensuremath{\hat{c}_4}}
\newcommand{\C}{\ensuremath{\mathbb C}}
\newcommand{\Z}{\ensuremath{\mathbb Z}}
\newcommand{\R}{\ensuremath{\mathbb R}}
\newcommand{\rp}{\ensuremath{\mathbb {RP}}}
\newcommand{\cp}{\ensuremath{\mathbb {CP}}}
\newcommand{\vac}{\ensuremath{|0\rangle}}
\newcommand{\vact}{\ensuremath{|00\rangle}                    }
\newcommand{\oc}{\ensuremath{\overline{c}}}

\newcommand{\Vol}{\textrm{Vol}}

\newcommand{\half}{\frac{1}{2}}
\begin{titlepage}
\bigskip
\def\thefootnote{\fnsymbol{footnote}}

\begin{center}
{\large {\bf
ABJ(M) and Fractional M2's with Fractional M2 Charge
  } }
\end{center}

\bigskip
\begin{center}
{\large  Jarah Evslin$^{1}$ and Stanislav Kuperstein$^{2}$}
\end{center}

\renewcommand{\thefootnote}{\arabic{footnote}}

\begin{center}

\vspace{1em}

{\em  {$^1$ SISSA,\\
Via Beirut 2-4,
I-34014, Trieste, Italy\\}} \texttt{evslin@sissa.it}

\vskip .4cm

{\em  { $^2$ Theoretische Natuurkunde,
Vrije Universiteit Brussel \\ and The International Solvay Institutes,
Pleinlaan 2,  B-1050 Brussels, Belgium \\}} \texttt{skuperst@vub.ac.be}

\end{center}

\vspace{.8cm}

\noindent
\begin{center} {\bf Abstract} \end{center}

\noindent

Recently Aharony, Bergman and Jafferis (ABJ) have argued that a 3$d$
$U(N+M)_k\times U(N)_{-k}$ Chern-Simons gauge theory may have a vacuum with
$\N=6$ supersymmetry only if $M \leqslant k$ and if a certain period of the $B$-field
in a IIA background is quantized.  We use a braneology argument to argue that
$\N=3$ supersymmetry may be preserved under the weaker condition that
$2Nk\geqslant M(M-k)$ with no restriction on the $B$-field.  IIB brane cartoons and
$11d$ supergravity solutions corresponding to $\N=3$ vacua that do not preserve
$\N=6$ supersymmetry are argued to represent cascading gauge theories,
generalizing the $\N=2$ Seiberg duality conjectured by Giveon and Kutasov.
While as usual the M2-brane charge runs as a result of the twisted Bianchi
identity for $*G_4$, the M5-brane charge running relies on the fact that it wraps a torsion homology cycle.





\vfill

\begin{flushleft}
{\today}
\end{flushleft}
\end{titlepage}

\hfill{}


\setcounter{footnote}{0}

\section{Introduction}

Cascading 4-dimensional gauge theories have been intensely studied over the
past decade.  In string theory there are two approaches to realizing these
theories, using Hanany-Witten brane cartoons \cite{HW} of D4-branes stretched between
NS5-branes and using branes wrapping cycles in a space which is topologically
$S^2\times S^3$ as in \cite{KN,KT,KS}.  In such constructions there are two RR charges,
corresponding to  the rank of one gauge group and to the difference between the
ranks.  The former runs while the latter is fixed.  The running of the former
may be seen \cite{Elitzur:1997fh} in the brane cartoon as the fact that a D4 stretched between two
NS5-branes wraps a compactified circle one less time each time that an
NS5-brane moves around the circle.  In the geometric realization the running
is the result of a twisted Bianchi identity for the improved field strength
$F_5$.  The Bianchi identity for the field strength $F_3$ is not twisted and
so the other charge does not run.

Cascading 3-dimensional Chern-Simons theories, as studied in \cite{GK,Vasilis1,Vasilis2,Alberto,Adi},
are different.  In these cases both charges run.  In the brane cartoon the
D3-branes are now extended between 5-branes with different charges, and so
when the 5-branes cross the Hanany-Witten effect leads to brane creation which
changes the second charge.  The geometric realization is a discrete quotient of $S^7$ and the two charges correspond to M2-branes and M5-branes.  The first charge, as in the 4-dimensional case, is
violated by a twisted Bianchi identity for the field $*G_4$.  As in the
4-dimensional case, the other Bianchi identity, that for $G_4$, is not
twisted.  However the M5-brane charge runs nonetheless because the M5-branes
wrap a 3-cycle which, unlike the $S^2$ in the 4-dimensional case,
represents a $\Z_k$ torsion homology class.  Therefore the number of M5-branes with this wrapping is only conserved modulo $k$.

This approach to fractional M2-brane charge is quite different from that
commonly found in the literature, as in our solutions $G_4$ is closed.  In the
case of the D3-brane charge flow in the Klebanov-Strassler model \cite{KS} $F_5=dC_4$ is also closed,
however the D3-brane charge is measured not by $F_5$, but by the exterior derivative of the improved
field strength
\beq
d(dC_4 + B_2 \wedge dC_2) = H_3 \wedge F_3 \neq 0.
\eeq
In the present case, we use a similar improved field strength for the running
of the M2-brane charge but no 11-dimensional analogue exists for the M5-brane
charge.  

However our solutions are topologically $S^1$ bundles, and after
dimensionally reducing the $S^1$ we obtain a non-closed 4-form field strength
in the remaining 10 dimensions.  Thus in the reduced theory one may define a
D4-brane charge to be the exterior derivative of this 4-form.   We define the
M5-brane charge on the $S^1$ to be equal to this charge in the reduced
theory. The T-dual brane cartoon suggests that this notion of charge agrees
with the rank of the gauge group.  This definition of charge via a
dimensionally-reduced theory is already familiar in torsional heterotic
compactifications on $T^2$ bundles over $K3$, where the tadpole condition
counts 5-branes wrapped on the topologically trivial $T^2$ and so is measured
by a dimensionally-reduced charge on the $K3$ base \cite{FuYau,BeckerFuYau,MeRuben}.  

The 3-dimensional gauge theories that we will consider in this note are also
different from 4-dimensional gauge theories in that the existence of
supersymmetric vacua depends on the choice of the gauge group.  In
\cite{ABJ} (ABJ) the authors showed that a necessary condition for $\N=6$
supersymmetry in a $U(N)_k\times U(N+M)_k$ Chern-Simons gauge theory is
$k\geqslant M$.  We will extend this result to argue that $\N=3$ supersymmetry
requires the weaker bound $2kN\geqslant M(M-k)$.  The two sides of this inequality
are suggestive of numbers of particles in a multiplet, for example the left
side may refer to $2k$ flavors of particles that transform in the fundamental
representation of $U(N)$, and the right hand side to adjoint $U(M)$ fields
minus those that are Higgsed by $k$ fields transforming in the fundamental of
$U(M)$.  At first it may seem counterintuitive that $k$ would lead to a
number of particles, as it is simply the Chern-Simons level.  However the
parity anomaly demonstrates that a Chern-Simons term at level $k$ has the same
contribution to the path integral as, for example, $k$ families of fermions
with infinite real masses.  Thus an alternate formulation of these theories
may exist where the Chern-Simons coupling is replaced by particles whose role
in the violation of the second charge may be understood.

The paper is organized as follows.
In Section \ref{typeIIBcartoon} we review the type IIB brane construction that we will use and derive the $\N=3$ S-rule bound, which is weaker than ABJ's $\N=6$ supersymmmetry condition.  We then repeat the arguments of \cite{GK} for a duality cascade. Section \ref{Mtheory} is devoted to the T-dual and the M-theory lift of the brane configuration. In particular we find that the distance between the 5-branes in type IIB corresponds to certain potentials in type IIA and M-theory.  In Section \ref{flus} we describe how the various brane charges are encoded in the fluxes in M-theory, arguing that the flow of the ranks in the cascade may be determined from the topology of the configuration, with the torsion third homology group responsible for the fact that both charges flow.  We conclude with some remarks on the possibility of finding a corresponding M-theory solution.

This morning the preprint \cite{Ofer} appeared which derives many of our results from a complimentary perspective.  The discussion of the IIB brane cartoons is quite similar, however while our note emphasizes the M-theory dual, the authors of \cite{Ofer} worked largely in type IIA.

\section{The type IIB brane cartoon}
\label{typeIIBcartoon}

In   \cite{9702202} the authors consider a configuration of intersecting
branes in type IIB string theory preserving 6 supercharges, and demonstrate
that the M-theory lift of its T-dual is a compactification on an 8-dimensional
hyper-K\"ahler manifold.  This construction has been exploited in
 \cite{ABJM} (ABJM), where it is argued that the (2+1)-dimensional 6 supercharge $U(N)\times U(N)$ 
Chern-Simons gauge theory has a $U(N)\times U(N)$ Chern-Simons theory with 12 supercharges as an infrared fixed point, and in fact describes a stack of M2-branes on the orbifold $\C^4/\Z_k$ (see \cite{Imamura:2008nn,Jafferis:2008qz,Fujita:2009xz} for various generalizations with less
supersymmetry).  
In \cite{ABJ} this construction was generalized to the case of a $U(N+M)\times U(N)$ gauge theory by including fractional M2-branes. 
These fractional M2-branes are rather subtle objects, as for example when $k=1$ there is no singularity 
in the geometry, and we will describe them in more detail below.  On the other hand they are quite easy 
to understand in the IIB brane cartoon.  And so we will begin by describing the configuration in type 
IIB and then in the next section carefully T-dualize and lift to M-theory.


\subsection{The $\N=3$ bound}


Consider type IIB string theory on $\R^{8,1}\times S^1$, where the circle direction is 
named $x^6$ and has period one.  
There is an NS5-brane extended along the directions $x^0$ through $x^5$ and at $x^6=x^7=x^8=x^9=0$.  
In the configurations of the above references there is also a ($k$,1) 5-brane, which is a bound 
state of $k$ D5-branes and a single NS5-brane.  
In the special case $k=1$ we may S-dualize the ($k$,1) 5-brane to a D5-brane.  
While this S-duality is not essential, 
it will make the following arguments somewhat simpler.  
The ($k$,1) 5-brane is placed diagonally with respect to the 
NS5-brane, extending again along $x^0$ through $x^5$ but now 
located at $x^7=kx^3$, $x^8=kx^4$ and $x^9=kx^5$.  We will 
name its position on the circle $x^6=y<1$.  

The 3-dimensional gauge theory lives on D3-branes which extend along 
the coordinates $x^0,\ x^1,\ x^2$ and $x^6$ with all 
other coordinates equal to zero.  Following   \cite{9702202}, we will draw a distinction between overlapping and intersecting branes.  The first are branes that are coincident but may move independently.  The latter are branes that are actually attached, so that it would at least require some energy, if not some charge conservation violation, to separate them.  While the branes in the above references are all overlapping, we will also consider branes which are intersecting.  

In particular, we consider $N$ D3-branes which wrap the entire $x^6$ circle.  These are the overlapping branes of \cite{ABJM}.  In addition we include $M$ D3-branes, which we will call fractional branes anticipating their T-duality although in type IIB they are ordinary D3-branes, which extend from $x^6=0$ to $x^6=y$.  These are in general intersecting branes, attached to the NS5-brane on one side and a $(k,1)$ 5-brane on the other.  The S-rule \cite{HW} limits how many of these intersecting D3-branes may exist in a supersymmetric configuration.  We will now determine this bound.

If $N=0$ the bound is just $M\leqslant k$ \cite{ABJ}, as one obtains by compactifying the results of \cite{BH,Ohta} on the $x^6$ circle
(see also \cite{Kitao1,Kitao2} for a related discussion of the S-rule for configurations of rotated branes).  The corresponding brane cartoon is identical to that of \cite{BH,Ohta} except that now the $x^6$ direction is compactified.  The physics on a space with a compactified direction is the same as that on the universal cover, but restricting to periodic configurations.  On the universal cover $x^6$ is replaced with an infinite line with periodically spaced and alternating NS5-branes and $(k,1)$ 5-branes, there is no compactified direction, and so we may use the S-rule of \cite{BH,Ohta} to conclude that the number of D3-branes stretched between any NS5-brane and $(k,1)$ 5-brane on the cover is at most equal to $k$.  

We may decompose the length of this D3-brane into a greatest integral part $j$
and a non integral part equal to $y$.  The integer $j$ is the number of
NS5-branes that it overlaps in the universal cover, or equivalently the number
of times that it wraps the $x^6$ circle.  In the compactified space the
world-volume theory of this brane contains a $U(j+1)\times U(j)$ gauge theory,
as it extends $j+1$ times between the NS5-brane and $(k,1)$ 5-brane and only
$j$ times around on the other side.  More generally, the S-rule tells us that
a stack of at most $k$ such D3-branes may exist in a supersymmetric
configuration, yielding a $U((j+1)k)\times U(jk)$ gauge theory. 

At this point one may object that the lift to the universal cover contains
more information, the number $j$, than was present in the original theory and
so is inequivalent.  However at any finite $g_s$ the D3-branes are described
by a BIon-like solution in which they are blown up into 2-spheres of 5-brane
which carry D3-brane charge.  The radius of this 2-sphere depends on $x^6$, 
and this dependence determines the number $j$.  Thus the number $j$
is not only evident in the universal cover, but also in the original
compactification at finite $g_s$.  Our S-rule may therefore be interpreted as
a bound on the number of tubes of D5-brane with each radius profile. 

\begin{figure}
\begin{center}
\includegraphics[width=0.6\textwidth]{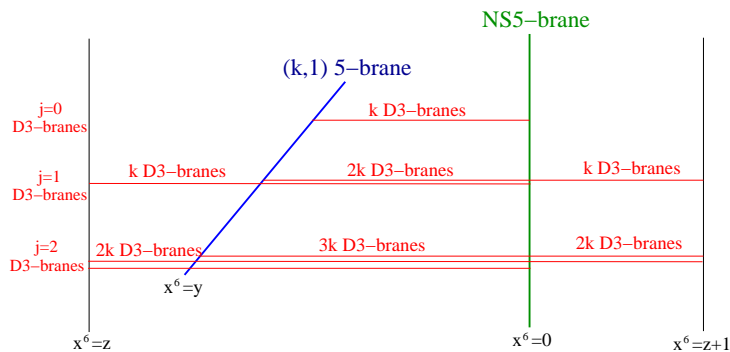}
\caption{This is a $\N=3$ supersymmetric brane cartoon with D3-branes corresponding to $j=0$, $j=1$ and $j=2$ extended between 5-branes.  The total gauge group is $U(3k)\times U(6k)$ which violates ABJ's $\N=6$ bound but satisfies the $\N=3$ bound of (\ref{n=3bound}).  It consists of $3k$ D3-branes which wrap the compactified $x^6$ circle various numbers of times.}
\label{n=3fig}
\end{center}
\end{figure}

This brings us to the question of just how large the difference 
$M$ between the ranks of the two gauge groups may be.  
In the above example the difference $M$ is equal to $k$, which satisfies ABJ's 
bound for $\N=6$ supersymmetry $M\leqslant k$. 
However combining two stacks of $k$ branes with different values of $j$ one easily violates the bound.  
For example in Fig.~\ref{n=3fig} we see an example with $j=1$, $2$ and $3$ all superimposed yielding 
a $U(3k)\times U(6k)$ gauge theory.  Now $M=3k>k$ and so ABJ's bound is not satisfied, 
the $\N=6$ supersymmetry is broken.  However the S-rule is satisfied and so not all of 
the supersymmetry is necessarily broken.  In fact, in a small coupling 
limit in which the 5-branes do not bend, $\N=3$ supersymmetry is preserved.  
This does not mean that the full 
$\N=3$ supersymmetry really is preserved at finite 
coupling, although at finite coupling the $SO(3)$ 
isometries corresponding to the SU(2) R-symmetry do appear to be 
preserved and so it does seem plausible that the full $\N=3$ 
supersymmetry is preserved in such cases.  However we 
will make the more modest claim, that when such combinations of $j$'s do 
not exist for a given $M$, that is to say 
when the S-rule is violated on the universal cover, then $\N=3$ supersymmetry is broken.  
This leaves us with a simple task, 
we need only determine for which values of $M$, $N$ and $k$ one can construct a brane 
cartoon whose universal cover 
satisfies the S-rule.

When $M=0$ there are no intersecting branes and so the S-rule places no constraint.  Instead we will seek an upper bound for $M$ at a given value of $N$, analogous to that of ABJ.  For a fixed value of $N$, one may obtain the largest value of $M$ by saturating the S-rule for the smallest values of $j$.  In other words, the largest value of $M$ is obtained by having $k$ D3-branes at $j=0$, another $k$ at $j=1$ and so on up to a maximum $j=i$ (for example, $i=2$ for the configuration on Fig.~\ref{n=3fig}). 
At each value of $j$ the D3-branes lead to a gauge group $U((j+1)k)\times U(jk)$.  Therefore the total gauge symmetry is
\beq
G=U \left(\frac{(i+1)(i+2)k}{2} \right)\times U \left( \frac{i(i+1)k}{2} \right)\hsp
N=\frac{i(i+1)k}{2}\hsp M=(i+1)k.
\eeq
Inverting this equation to find $N$ one obtains the lower bound on $N$ for a given $k$ and $M$
\beq
N\geqslant \frac{M(M-k)}{2k}. \label{n=3bound}
\eeq
Notice that when the ABJ $\N=6$ bound is satisfied  the right hand side of the $\N=3$ bound (\ref{n=3bound}) is non positive, therefore the $\N=3$ bound is trivially satisfied.



\subsection{RG flow}

As in the case of 4-dimensional gauge theories, the renormalization group flow of the coupling constant may be read from the brane cartoon.  The inverse coupling squared at a given energy scale is just the $x^6$ separation of the 5-branes at a given position in the transverse directions.  Therefore the RG flow of the coupling constant is determined by the change in separation, or more intuitively, by the curvature of the 5-branes resulting from the fact that they are pulled by the D3-branes.

In particular a conformal $\N=6$ theory may exist only if the 5-branes are not curved.  In general, this means that there must be the same number of D3-branes pulling on each 5-brane from each direction.  This is the case for example when $M=0$, in which case the D3-branes are merely overlapping the 5-branes and so exert no force.  Any nonzero value of $M$ will imply that there are $M$ more D3-branes pulling each 5-brane in one direction, and so generically will cause the 5-branes to bend.  There is only one exception to this heuristic rule, if the D3-branes have zero length.  In other words, if the two 5-branes are at the same $x^6$ position, so that $y=0$, then the 3-branes exert no net force.

If one separates the 5-branes by a small distance $y>0$, then there will 
be $M$ D3-branes connecting them. Solving the 3-dimensional\footnote{The $3d$ space is spanned by $(x_3,x_4,x_5)$ 
or alternatively by $(x_7,x_8,x_9)$.} 
Laplace equation for the transverse positions of the 5-branes, 
one finds that the D3-branes bend the D5-branes only very close to the origin, and only by a small amount. In the limit 
$y\mapsto 0$ the 5-branes are coincident as above. One may also 
consider a small displacement in the other direction $y<0$. As the 
5-branes cross, the Hanany-Witten transition implies that there are now 
$k-M$ D3-branes. Again one may take the limit $y\mapsto 0$, and one 
arrives at the same coincident 5-brane configuration, but now with $k-M$ 
infinitesimal D3-branes instead of $M$. The assertion that these two 
configurations, obtained by taking the limit as $y$ goes to zero from 
above and below, must describe the same physics is the 
\beq 
M \longrightarrow k-M 
\eeq 
duality postulated by ABJ for $y=0$.
However the $\N=6$ supersymmetry only exists when the 5-branes do not bend 
(recall that the $\N=6$ supersymmetry imposes conformality), which implies $y=0$.  
In the next section we will see that in type IIA this corresponds also to a quantized $B$-field, 
explaining the quantization of the $B$-field observed by ABJ.

If the $\N=3$ theory is not conformal, ABJM's IR $\N=6$ fixed point does not exist.  Instead as one flows into the IR, corresponding to moving towards the origin in the transverse directions along the branes, the 5-branes appear to approach and cross.  This is quite similar to the duality cascade braneology of   \cite{KS}, except that now the Hanany-Witten transition implies an additional shift of $k$, as noted in \cite{GK,ABJ}.  In the covering space these steps are quite simple, each value of $j$ simply decreases by one and all $k$ of the $j=0$ D3-branes disappear.  More generally if there are $M$ $j=0$ branes, they are replaced by $k-M$.  In general one arrives at the duality conjectured for the $\N=2$ case in   \cite{GK}
\beq
U(N)_k\times U(N+M)_{-k}\longrightarrow U(N)_{-k}\times U(N+k-M)_{k}. \label{passo}
\eeq  

\begin{figure}
\begin{center}
\includegraphics[width=0.6\textwidth]{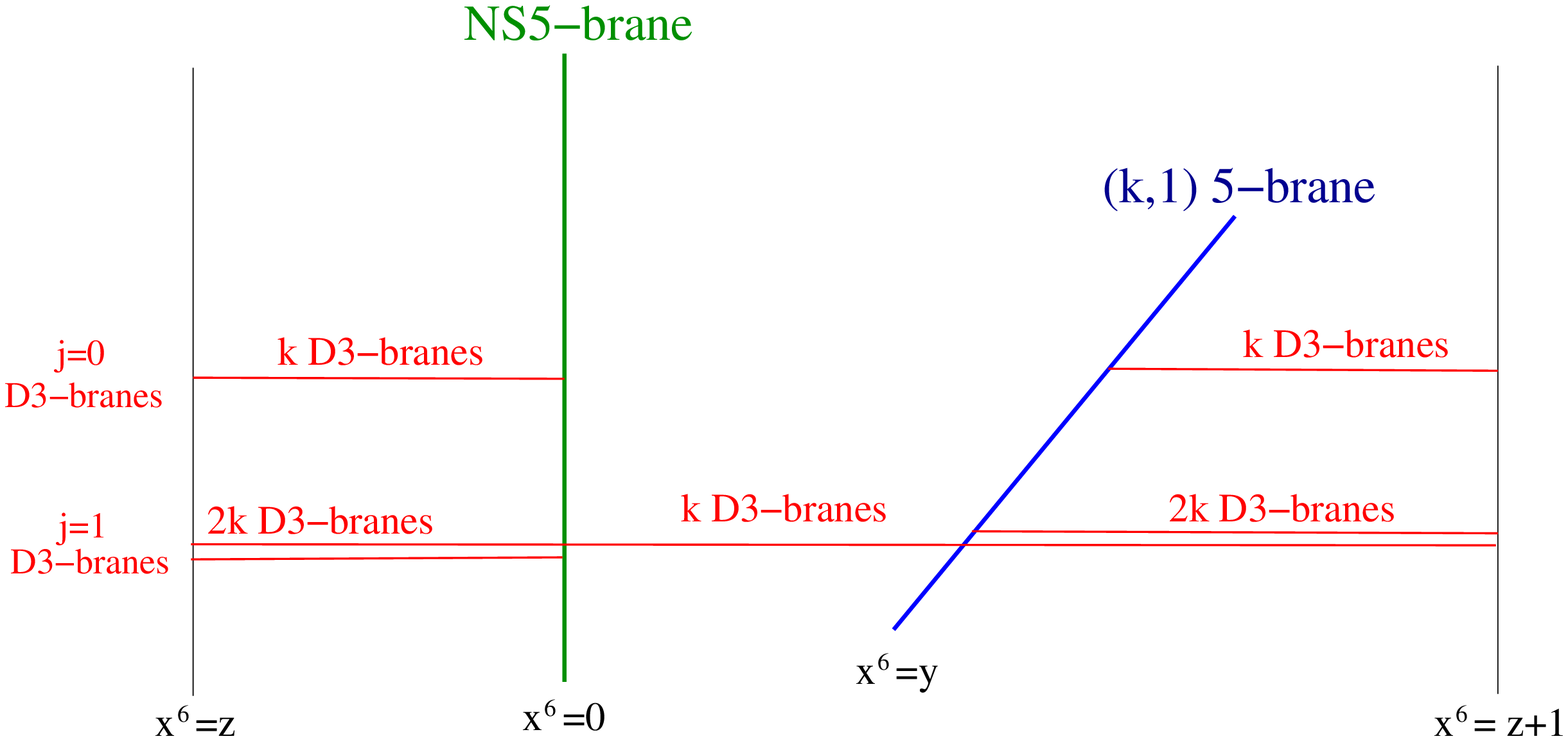}
\caption{This is a $\N=3$ supersymmetric brane cartoon with D3-branes corresponding to $j=0$ and $j=1$ extended between 5-branes.  The total gauge group is $U(2k)\times U(k)$ which violates ABJ's $\N=6$ bound but satisfies the $\N=3$ bound of (\ref{n=3bound}).  It consists of $2k$ D3-branes which wrap the compactified $x^6$ circle various numbers of times.}
\label{n=3dopo}
\end{center}
\end{figure}

For example consider the the $U(6k)\times U(3k)$ configuration of Figure~\ref{n=3fig}, corresponding to $N=M=3k$.  After one step this becomes the $U(3k)\times U(k)$ configuration of Figure~\ref{n=3dopo}.

Notice that the $\N=3$ condition (\ref{n=3bound}) is invariant under the step (\ref{passo}).  This implies that the conjecture that (\ref{n=3bound}) is a necessary and sufficient condition for the existence of an $\N=3$ vacuum is compatible with the Seiberg-like duality (\ref{passo}) of   \cite{GK}.

\section{Duality to M-theory}
\label{Mtheory}

\subsection{T-duality to IIA}
Now we are ready to T-dualize along $x^6$ to obtain a type IIA configuration.
For simplicity we first consider $k=1$ and S-dualize the brane cartoon so that
there is a single NS5-brane and a single D5-brane.  Later we will describe how
the situation changes for general $k$ and without the S-duality.  The case
$y=0$ was already considered in   \cite{9702202}, more precisely they
smeared all branes in the $x^6$ direction and so $y$ was not defined.  We will also refer to the new circle in the type IIA compactification as $x^6$.  We set $\alpha\p$ to one, and so $x^6$ still has periodicity equal to one.  

The NS5-brane is T-dual to a KK monopole with respect to the $x^6$ circle,
located again at $x^7=x^8=x^9=0$.  The D5-brane is T-dual to a D6-brane
located again at $x^7=x^3$, $x^8=x^4$ and $x^9=x^5$ and wrapping the $x^6$
circle.  The $x^6$ circle degenerates at the origin in the $x^7$,
$x^8$ and $x^9$ directions, and so one may wish, when possible, to slightly displace the various D6-branes from this point to better visualize the geometry in what follows.  The $x^6$ coordinate $y$ of the D5-brane becomes a Wilson line on the D6-brane along the $x^6$ direction $A=y dx^6$.  

\begin{figure}
\begin{center}
\includegraphics[width=0.6\textwidth]{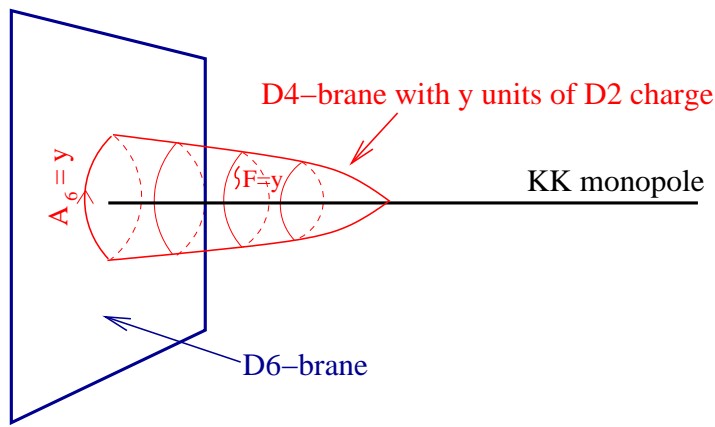}
\caption{In the case $k=1$ the S-dual type IIA configuration consists of a D6-brane which intersects a KK-monopole.  In general there are fractional D2-branes at the intersection.  In this picture the fractional branes have been moved off of the singularity, becoming D4-branes.  Their world-volume gauge theories have nontrivial Wilson lines, which implies by Stokes' theorem that they have nontrivial gauge field strengths, yielding fractional D2-brane charge.}
\label{IIA}
\end{center}
\end{figure}

The $N$ D3-branes are now D2-branes located at the origin and extending along $x^0$ through $x^2$.  The $M$ fractional D3-branes are also mapped to the origin, where one would expect them to carry $My$ units of D2-brane charge.  To see that this is indeed the case, we may break the supersymmetry by slightly displacing the D2-branes from the intersection of the KK monopole and D6-brane as in Fig.~\ref{IIA}.  Now the fractional branes are D4-branes which extend from the KK monopole to the D6-brane.  They wrap the $x^6$ circle, which degenerates at the KK monopole but not at the D6-brane.  Therefore they are cylinders which shrink at one end, and so they have disc topology.  Their boundary is the $x^6$ circle on the D6-brane.  

At finite orders in $g_s$ the boundary is really not a sharp angle, but instead a BIon configuration where the D4-brane grows into a sphere of the D6-brane which continuously merges with the D6. This means that the D4-branes share the D6-brane's Wilson loop $A_6=y$.  However unlike the D6-branes, the D4-branes are discs and so this Wilson loop is supported on a contractible cycle.  One may therefore use Stokes' theorem to calculate the world volume field strength inside of the D4-brane
\beq
\int_{D^2}F = \int_{D^2} dA = \int_{x^6} A = y.
\eeq
This produces a contribution to the D4-brane Wess-Zumino terms
\beq
S_{WZ}\supset\int_{D^2\times\R^{2,1}} C_3 \wedge F = y \int_{\R^{2,1}}C_3
\eeq
therefore identifying the fractional D2-brane charge of each D4-brane as $y$, in line with the expectations from the IIB side.  In particular, if one undoes the above displacement of the D6-brane, one will find that the fractional D2-branes have a tension which is only $y$ times the tension of the other D2-branes.

What happens with different values of $k$?  Now the D6-brane will necessarily
be a superposition of KK monopoles, and so naively the $x^6$ circle always
shrinks on top of the D6-brane and this argument does not apply.  However in
M-theory one sees that it is not actually the $x^6$ circle that shrinks, but
rather a linear combination of the $x^6$ and $x^{10}$ circles.  The D4-branes will lift to M5-branes which wrap both, and so they will still have a nontrivial boundary on the non vanishing circle which allows the fractional brane argument to work.  We will make this more precise in the next subsection.

\subsection{M-theory lift}

In the M-theory lift, all of our original 5-branes have disappeared.  The
NS5-brane already disappeared in type IIA, leaving a Kaluza-Klein monopole with
respect to the $x^6$ circle.  Now the D5-branes, which was T-dual to a
D6-brane also disappears, leaving a Kaluza-Klein monopole for the $x^{10}$
M-theory circle.  For $k>1$ with no S-duality it yields a KK monopole for the
$x^6+kx^{10}$ circle.  These two monopoles may simply be superimposed \cite{9702202}, yielding a hyper-K\"ahler space that preserves at least 6 supercharges.  Near the origin, this space in fact preserves twice as many supercharges \cite{ABJM} and is simply $\C^4/\Z_k$, which in the $k=1$ case is $\C^4$. 

However when $y\neq 0$ the M-theory lift of \cite{ABJ,9702202,ABJM} is
somewhat altered.  This is because the D6-brane carries a nontrivial Wilson
loop.  In   \cite{Sparks} the author argued that the world-volume gauge
field strength of a D6-brane is equal to the integral of the M-theory 4-form
field strength over a 4-cycle which is a circle-valued family of 3-cycles
which are M-theory lifts from discs ending on the D6-brane.  We will see below
that the same correspondence holds for the connections.  In other words, the
D6-brane world-volume Wilson line lifts to an M-theory 3-form connection $C_3$
on a 3-cycle which is the M-theory lift of a 2-disc bounded by the D6-brane.
Momentarily we will see that this $C_3$ flux is also required by Dirac quantization.



So what is the lift of our fractional branes?  They were D4-branes extended
from the $x^6$ KK-monopole to the D6-brane and wrapping the $x^6$ circle.
Therefore they must now be M5-branes wrapping the $x^6-x^{10}$ torus and
extending between an $x^6$ KK-monopole and an $x^6+kx^{10}$ KK-monopole.  The
2-torus is fibered over the line interval extending between the two monopoles,
which is nontrivial when slightly displaced from the origin.  A 2-torus fibered over a line interval, with two circles degenerating on both ends, describes the lens space $L_{k,1}=S^3/\Z_k$ where $k$ is the index of subgroup of the first homology of the torus which is spanned by the degenerating circles.  This is the same lens space as is wrapped in the large $N$ dual $AdS_4\times S^7/\Z_k$ of   \cite{ABJ}.  This $k$ is equal to the Chern-Simons level $k$ above, since the circles correspond to the elements $(k,1)$ and $(0,1)$ which indeed generate an index $k$ subgroup of $\Z_2$.  

If one compactifies this lens space on a circle along the $\Z_k$ direction then one arrives at a D4-brane on a 2-sphere with $k$ units of $F_2$ flux, and so the Wess-Zumino term implies
\beq
S \supset \int F_2\wedge A\wedge dA = k \int A\wedge dA
\eeq
that the fractional branes yield 3-dimensional Chern-Simons theories at level $k$ as desired \cite{ABJ,AcharyaVafa}.

The lift of the 2-form gauge field strength $F$ on the D4-brane is a 3-form self-dual field strength $T$ on the M5-brane, which will be equal to $T=F\wedge dx^{10}$.  Therefore one may use the M5-brane Wess-Zumino term to determine the M2-brane charge
\beq
S\supset \int_{S^3/\Z_k\times\R^{2,1}} C_3 \wedge T =\int_{S^3/\Z_k\times\R^{2,1}} C_3\wedge F\wedge dx^{10} 
  = \int_{D^2\times\R^{2,1}} C_3 \wedge F = y \int_{\R^{2,1}}C_3.
\eeq
Therefore the M5-branes indeed each carry $y$ units of M2-brane charge.  

This may appear surprising if one thinks that the 3-form $T$ needs to be
quantized.  However, as is the case with the gauge field strength on a
D-brane, the gauge-invariant 3-form field strength $T$ is the sum of a closed,
quantized piece $dA_2$ and a pullback of the bulk 3-form $C_3$.  Therefore we
learn that the pullback of the M-theory 3-form $C_3$ to this lens space is
equal to $y$.  Therefore the Dirac quantization of $dA_2$, the M2-brane
Page charge, implies that $\int C_3=y$ as claimed above.

Dimensionally reducing to type IIA, this implies that the integral of the 
$B$-field is equal to $y$. In particular, in the $\N=6$ case in which the IIB 
5-branes do not bend, 
$y$ is an integer and so the $B$-field is quantized exactly as was advertised in the previous section.


In summary, the parameter $y$, which was necessarily equal to 0 in the $\N=6$ case, determines the fractional brane charge.  It is manifested differently in the various theories, as is summarized in Table~\ref{ytab}.

\noindent
\begin{table}
\begin{tabular}{c|c}
\bf{Theory}&\bf{Manifestation of $y$}\\\hline
IIB Brane Cartoon&Distance between 5-branes/inverse gauge coupling squared\\
IIA D6-brane theory&Wilson loop on $x^6$ circle\\
IIA D4-brane theory&Gauge field strength on disc/fractional D2 charge\\
M-theory M5-brane theory&$T$ field strength on lens space/fractional M2 charge\\
M-theory bulk fields&$C_3$ gauge connection on lens space\\
\end{tabular}
\caption{{\textit{Our construction generalizes those in the literature by introducing a parameter $y$.  In this table we describe how $y$ appears in the various dual descriptions.}}} \label{ytab}
\end{table}

\subsection{Seiberg dualities}

The tension of the fractional D3-branes in type IIA will bend the 5-branes, leading to a renormalization group flow of the $U(N+M)\times U(N)$ gauge theory which may well change the ranks of the gauge groups.  In particular, in   \cite{ABJ} the authors have found a single such duality\footnote{
See also \cite{Alberto,Adi} for a recent discussion of Seiberg dualities in $3d$ theories.}, which essentially interchanges $M$ and $k-M$.  

The running is somewhat different from the 4-dimensional running of Witten \cite{MQCD} and Klebanov-Strassler \cite{KS}.  Those field theories were described by brane cartoons in type IIA, in which the gauge theory lives on a D4-brane stretched between two NS5's.  The length of the D4, which is the separation of the NS5's, is identified with the inverse gauge coupling squared $1/g_{YM}^2$.  As the intersection of the D4 and NS5's is co-dimension two on the NS5's, this separation solves a 2-dimensional Laplace equation and so is logarithmic, leading to the usually logarithmic RG flow in 4-dimensions.  In particular at sufficiently high energy the logarithm is arbitrarily large and so the NS5-branes are arbitrarily distant.  When the separation direction is compactified this implies that the branes cross an infinite number of times, yielding a solution with an infinite cascade.


As we have reviewed, 3-dimensional gauge theories are modeled, following   \cite{HW}, by brane cartoons in type IIB with D3-branes stretching between 5-branes.  The endpoints are now co-dimension 3 and so the RG flow is captured by the constant plus $1/E$ solution of the 3-dimensional Laplace equation.  In particular there is now a maximum distance between the 5-branes, corresponding to the fact that the gauge coupling is relevant in 3-dimensions and so the 1-loop RG flow is asymptotically unimportant in the ultraviolet.  Compactifying this direction one finds that the UV completion of the cascade contains a finite number of steps.  At each step the new gauge group begins with the new gauge coupling. When the energy scale exceeds gauge coupling squared, the cascade ends.


\section{The cascade from M-theory} \label{flus}

The results of the present note allow one to see such transitions directly in the M-theory configuration.  The M2-charge of $M$ M5-branes is
\beq
Q_2=M\int_{S^3/\Z_k}C_3.
\eeq
Increasing $C_3$ by one, corresponding to bringing a 5-brane all of the way around the IIB circle, then increases the M2-charge by $M$.  What about the shift by $k$?  

To describe this, we will need to be a bit more explicit about our choice of
fields.  Topologically our 11-dimensional spacetime is $\R^{2,1}$ times a cone
over $S^7/\Z_k$.  This quotient is an $S^1$ fibered over $\cp^3$ with Chern class $c_1=k$.  In particular the third homology group is
\beq
\H_3(S^7/\Z_k,\Z)=\Z_k
\eeq
which is generated by the lens space $S^3/\Z_k$.  In other words there exists a 4-chain $\Sigma_k$ whose boundary is homologous to $k$ copies of $S^3/\Z_k$.

Let $r$ be the radial direction along the cone.  Then the $C_3$ flux may be decomposed into a component on the $S^3/\Z_k$, on $\R^{2,1}$ and on $\Sigma_4$.  We will choose this $S^3/\Z_k$ so that the integral of $C_3$ over $S^3/\Z_k$ is just the T-dual of the distance between the 5-branes
\beq
\frac{1}{g_{YM}^2}=\int_{S^3/\Z_k}C_3=y\sim a-\frac{b}{E}
\eeq
where $a$ and $b$ are constants for each gauge group and $E$ is the energy.

Let $G_4=dC_3$ be the 4-form field strength.  Then the Bianchi identity
\beq
d \star_{11} G_4 = \half G_4\wedge G_4 \label{bian}
\eeq
may be used to determine the running of the M2-brane charge $Q_{\rm{M2}}$, just as the Bianchi identity for the 5-form determines the running of the D3-brane charge in the Klebanov-Strassler cascade.  The M2-brane charge at radius less than $r_0$, corresponding to the contribution to the ranks of both gauge groups at a fixed energy, is given by the integral of $d*G_4$ up to that value of $r$, which by (\ref{bian}) is the integral of $G_4\wedge G_4$.  By Stokes' theorem this may be re-expressed as an integral on the base
\beq
\int_{r<r_0} d \star_{11} G_4 = \half \int_{r<r_0}G_4\wedge G_4 = 
   \half \int_{S^7/\Z_k} C_3\wedge G_4 = \half  \left( \int_{S^3/\Z_k}C_3 \right) \left( \int_{\Sigma_4}G_4 \right).
\eeq
Note in particular that it depends on $\int_{S^3/\Z_k}C_3$ which we identified as the distance between the 5-branes in type IIB.  However to determine the charge we will also need to calculate $\int_{\Sigma_4}G_4$.

What about the M5-brane charge $Q_{\rm{M5}}$?  This is given by $dG_4$, which by Stokes' theorem we may write as the integral of $G_4$ on the $\Sigma_4$.  Summarizing
\beq
Q_{\rm{M5}}=\int_{\Sigma_4}G_4\hsp
Q_{\rm{M2}}=\half\left( \int_{S^3/\Z_k}C_3 \right) \left( \int_{\Sigma_4}G_4 \right) \hsp
\int_{S^3/\Z_k}C_3\sim a-\frac{b}{E}. \label{cariche}
\eeq
At first glance this is not in agreement with our expectations for the Seiberg duality.  The Hanany-Witten transition in type IIB should change the difference in the ranks of the gauge groups, which is the number of fractional branes $Q_{\rm{M5}}$.  However we have identified this charge as the integral of a closed form, which only depends on the homology class of the cycle on which it is wrapped and so seems independent of $r$.

The resolution to this problem is that this $\Sigma_4$ is not a closed submanifold of $S^7/\Z_k$ as
\beq
\partial \Sigma_4=kS^3/\Z_k.
\eeq
One may then wonder whether such an integral over a manifold with boundary may
measure a RR charge.  It can.  In fact such terms are familiar in the
heterotic tadpole condition for heterotic torsional compactifications on torus
bundles over $K3$, in these cases the K3 is not a submanifold but the tadpole
nonetheless is an integral over K3 which measures the charges of 5-branes
wrapped on the fiber \cite{FuYau,BeckerFuYau,MeRuben}.


Now we are now ready to calculate $\int_{\Sigma_4}G_4$.  By Stokes' theorem it is
\beq
Q_{\rm{M5}}=\int_{\Sigma_4}G_4=\int_{\partial \Sigma_4}C_3=k\int_{S^3/\Z_k}C_3=k  \left( a-\frac{b}{E} \right).
\eeq
This gives us the RG flows of the charges.  The M5-brane charge is linear in
$1/E$ and so by (\ref{cariche}) the M2-brane charge is quadratic with a
second derivative which is twice the slope of the M5-brane charge\footnote{The $E = \infty$ version of this result 
has already appeared in formula (3.14) of \cite{Oren} with $a=l/k$, where $l=M$ in the notations of \cite{Oren} 
(notice that with this identification $Q_{\rm{M5}}=M$ for $E = \infty$).  We are grateful to Oren Bergman for bringing 
this point to our attention as well as for correcting a numerical mistake in the earlier version of the paper.}
\beq
Q_{\rm{M2}}=\half k \left( \int_{S^3/\Z_k}C_3 \right)^2=\half k\left( a-\frac{b}{E} \right)^2.
\eeq
This is precisely what we expect from the cascade based on the T-dual IIB brane cartoons, as the difference in M5 charge at each step changes the step size of the M2-brane charge.  

More concretely, after $n$ steps we expect the M5-brane charge to change by $kn$ and the M2-brane charge to change by $n$ times the original M5-brane charge plus roughly $kn^2/2$.  Therefore the change in M2-brane charge is roughly the square of the change in M5-brane charge divided by $k$.  Here $n$ is the change in the integral of $C_3$ over the $S^3/\Z_k$.

This result is independent of the energy scales at which the transitions occur and of the deformations in the metric caused by the backreaction of the fluxes.  It is a consequence only of the torsion third homology group of $S^7/\Z_k$. 

\section{Towards a cascading SUGRA solution}

In this short section we make some comments on a potential $11d$ supergravity
solution for a cascading $3d$ gauge theory.  We begin with the usual M2-brane Ansatz
\begin{equation}
d s_{11}^2 = H^{-2/3} d x_\mu d x^\mu + H^{1/3} d s_{M_8}^2 
\qquad \textrm{and} \qquad
G_4 = d^3 x \wedge d H^{-1} + m L_{4},
\end{equation}
where the function $H$ depends only on the $M_8$ coordinates, likely on only
the radial direction $R$, and $L_{4}$ has no space-time legs.  Here $ds_{M_8}^2$ is the metric of Ref.~\cite{9702202}.
The $4$-form EOM is
\begin{equation}
d \star_{11} G_4 = \half G_4 \wedge G_4.
\end{equation}
In terms of $H$ and $L_{4}$ this equation implies that
\begin{equation}
\label{Leq}
\Box H = -\frac{1}{48} m^2 L_{4}^2
\qquad \textrm{and} \qquad
d  \left( \ln H \right) \wedge \left(  \star_8 L_{4} - L_{4} \right) = d \star_8 L_{4} .
\end{equation}

Notice that the last equation can be easily solved by a \emph{closed} and \emph{self-dual} $4$-form $L_{4}$.
More precisely assuming that $L_{4}$ is self-dual the above equation implies
also that it is closed.   

Let $\sigma_3$ be the top form on the $S^3/\Z_k$ submanifold, and let $f(r)$ be a
function of the radial direction such that
\beq
L_4=d(f(r)\sigma_3)=\star_8 L_4.
\eeq
Then the Bianchi identity will be satisfied.  Self-duality is likely to imply
$\N=2$ supersymmetry.  However it is quite possible that cascading solutions
exist which preserve $\N=3$.  In this case, one may re-express the self-duality
condition in terms of contractions with the K\"ahler form.  One then needs to
impose the condition with respect to two distinct K\"ahler forms to obtain
$\N=3$ supersymmetry.




\section*{Acknowledgments}

It is a pleasure to thank Oren Bergman, Matteo Bertolini, Carlos Nunez, Alberto Mariotti, Ami Hanany, Adi Armoni, Dario Martelli, Daniel Arean Fraga and especially Daniel Persson for useful conversations.



\begin{thebibliography}{23}



\bibitem{HW}
  A.~Hanany and E.~Witten,
  {\it Type IIB superstrings, BPS monopoles, and three-dimensional gauge
  dynamics},
  [arXiv:hep-th/9611230].


\bibitem{KN}
I.~R.~Klebanov and N.~A.~Nekrasov,
{\it Gravity duals of fractional branes and logarithmic RG flow}, [arXiv:hep-th/9911096].



\bibitem{KT}
I.~R.~Klebanov and A.~A.~Tseytlin,
{\it Gravity Duals of Supersymmetric SU(N) x SU(N+M) Gauge Theories}, [arXiv:hep-th/0002159].


\bibitem{KS}
I.~R.~Klebanov and M.~J.~Strassler,
{\it Supergravity and a confining gauge theory: Duality cascades and $\xi$SB
resolution of naked singularities}, [arXiv:hep-th/0007191].



\bibitem{Elitzur:1997fh}
  S.~Elitzur, A.~Giveon and D.~Kutasov,
  {\it Branes and N = 1 duality in string theory},
  [arXiv:hep-th/9702014].



\bibitem{GK}
  A.~Giveon and D.~Kutasov,
 {\it Seiberg Duality in Chern-Simons Theory},
  [arXiv:0808.0360].




\bibitem{Vasilis1}
  V.~Niarchos,
  {\it Seiberg Duality in Chern-Simons Theories with Fundamental and Adjoint Matter},
  [arXiv:0808.2771].

\bibitem{Vasilis2}
  V.~Niarchos,
  {\it R-charges, Chiral Rings and RG Flows in Supersymmetric Chern-Simons-Matter Theories},
  [arXiv:0903.0435].





\bibitem{Alberto}
  A.~Amariti, D.~Forcella, L.~Girardello and A.~Mariotti,
 {\it 3D Seiberg-like Dualities and M2 Branes},
  [arXiv:0903.3222].


\bibitem{Adi}
  A.~Armoni, A.~Giveon, D.~Israel and V.~Niarchos,
 {\it Brane Dynamics and 3D Seiberg Duality on the Domain Walls of 4D N=1 SYM},
  [arXiv:0905.3195].







\bibitem{FuYau}
J.~X.~Fu and S.~T.~Yau,
{\it Existence of supersymmetric Hermitian metrics with torsion on non-K\"ahler
manifolds}, [arXiv:hep-th/0509028].



\bibitem{BeckerFuYau}
K.~Becker, M.~Becker, J.~X.~Fu, L.~S.~Tseng and S.~T.~Yau,
{\it Anomaly cancellation and smooth non-K\"ahler solutions in heterotic string
theory}, [arXiv:hep-th/0604137].



\bibitem{MeRuben}
J.~Evslin and R.~Minasian,
{\it Topology change from (heterotic) Narain T-duality}, [arXiv:0811.3866].




\bibitem{ABJ}
  O.~Aharony, O.~Bergman and D.~L.~Jafferis,
  {\it Fractional M2-branes},
  [arXiv:0807.4924].




\bibitem{Ofer}
  O.~Aharony, A.~Hashimoto, S.~Hirano and P.~Ouyang,
  {\it D-brane Charges in Gravitational Duals of 2+1 Dimensional Gauge Theories
  and Duality Cascades},
  [arXiv:0906.2390].
  




\bibitem{9702202}
  J.~P.~Gauntlett, G.~W.~Gibbons, G.~Papadopoulos and P.~K.~Townsend,
 {\it Hyper-Kaehler manifolds and multiply intersecting branes},
  [arXiv:hep-th/9702202].

\bibitem{ABJM}
  O.~Aharony, O.~Bergman, D.~L.~Jafferis and J.~Maldacena,
  {\it N=6 superconformal Chern-Simons-matter theories, M2-branes and their
  gravity duals},
  [arXiv:0806.1218].



\bibitem{Imamura:2008nn}
  Y.~Imamura and K.~Kimura,
  {\it On the moduli space of elliptic Maxwell-Chern-Simons theories},
  [arXiv:0806.3727].


\bibitem{Jafferis:2008qz}
  D.~L.~Jafferis and A.~Tomasiello,
  {\it A simple class of N=3 gauge/gravity duals},
  [arXiv:0808.0864].
  
  
\bibitem{Fujita:2009xz}
  M.~Fujita and T.~S.~Tai,
  {\it Eschenburg space as gravity dual of flavored N=4 Chern-Simons-matter theory},
  [arXiv:0906.0253].




\bibitem{BH}
  O.~Bergman, A.~Hanany, A.~Karch and B.~Kol,
  {\it Branes and supersymmetry breaking in 3D gauge theories},
  [arXiv:hep-th/9908075].
 
\bibitem{Ohta}
  K.~Ohta,
  {\it Supersymmetric index and s-rule for type IIB branes},
  [arXiv:hep-th/9908120].




\bibitem{Kitao1}
  T.~Kitao, N.~Ohta and J.~G.~Zhou,
  {\it Fermionic zero mode and string creation between D4-branes at angles},
  [arXiv:hep-th/9801135].


\bibitem{Kitao2}
  T.~Kitao, K.~Ohta and N.~Ohta,
  {\it Three-dimensional gauge dynamics from brane configurations with (p,q)-fivebrane},
  [arXiv:hep-th/9808111].






\bibitem{Sparks}
 J.~Sparks,
  {\it Global worldsheet anomalies from M-theory},
  [arXiv:hep-th/0310147].
  




\bibitem{AcharyaVafa}
  B.~S.~Acharya and C.~Vafa,
  {\it On domain walls of N = 1 supersymmetric Yang-Mills in four dimensions},
  [arXiv:hep-th/0103011].
 

\bibitem{MQCD}
E.~Witten,
{\it Solutions of four-dimensional field theories via M-theory},
[arXiv:hep-th/9703166]



\bibitem{Oren}
  O.~Bergman and S.~Hirano,
  {\it Anomalous radius shift in AdS(4)/CFT(3)},
  [arXiv:0902.1743].
  
  

\end{thebibliography}
\end{document}